\documentclass{IEEEtran}
\usepackage{cite}
\usepackage{amsmath,amssymb,amsfonts}
\usepackage{algorithmic}
\usepackage{graphicx}
\usepackage{textcomp}
\def\BibTeX{{\rm B\kern-.05em{\sc i\kern-.025em b}\kern-.08em
    T\kern-.1667em\lower.7ex\hbox{E}\kern-.125emX}}

\usepackage{pdfpages}
\usepackage{threeparttable} %ÖÆ×÷ÈýÏß±í¸ñ
\usepackage{multirow} %ºÏ²¢¶àÐÐµ¥Ôª¸ñµÄºê°ü
\usepackage{booktabs}
\usepackage{enumerate}
\usepackage{algorithm}
\usepackage{algorithmic}
\usepackage{epsfig}
\usepackage{multicol} % Used for the two-column layout of the document
\usepackage{stfloats}

\newtheorem{theorem}{Theorem}
\newtheorem{proposition}{Proposition}
\newtheorem{lemma}{Lemma}

\newtheorem{remark}{Remark}
\newtheorem{definition}{Definition}
\newtheorem{assumption}{Assumption}

\makeatletter

\newcommand{\Rmnum}[1]{\expandafter\@slowromancap\romannumeral #1@}
\usepackage{setspace}                                    % printed Automatica style}
\begin{document}
\title{Distributed  State Estimation for Discrete-Time  Linear  Systems  over Directed Graphs: A Measurement Perspective}
\author{Xiaoxu Lyu, Guanghui Wen, Yuezu Lv, Zhisheng Duan, and Ling Shi

\thanks{
Xiaoxu Lyu and Ling Shi are with the Department of Electronic and Computer Engineering, The Hong Kong University of Science and Technology, Hong Kong, China. Ling Shi is also with the Department of Chemical and
Biological Engineering, The Hong Kong University of Science and Technology,
Hong Kong, China (e-mail: eelyuxiaoxu@ust.hk; eesling@ust.hk).}

\thanks{Guanghui Wen is with  the School of Automation, Southeast University, Nanjing 210096, China (e-mail: wenguanghui@gmail.com).}

\thanks{Yuezu Lv is with MIIT Key Laboratory of Complex-field Intelligent Sensing, Beijing Institute of Technology, Beijing 100081, China (e-mail: yzlv@bit.edu.cn).}

\thanks{Zhisheng Duan is with the School of Advanced Manufacturing and Robotics, Peking University, Beijing 100871, China (e-mail: duanzs@pku.edu.cn).}

}

\maketitle

\begin{abstract}
This paper proposes a novel   consensus-based  distributed filter over directed graphs under the collectively observability condition.  The distributed filter  is designed  using  an augmented leader-following  information  fusion strategy, and  the gain parameter is   determined    exclusively using   local information.
Additionally,  the lower bound of the fusion step number  is derived  to ensure  that  the estimation error covariance remains  uniformly upper-bounded. Furthermore, the lower bounds for  the convergence rates of the steady-state
performance gap between the proposed filter  and the centralized filter are provided  as the fusion step number  approaches  infinity.
The analysis demonstrates  that the convergence rate is  at least  as fast  as exponential convergence, provided  the communication topology satisfies  the  spectral norm condition.
 Finally,  the theoretical results are validated  through     two simulation examples.
\end{abstract}

\begin{IEEEkeywords}
Distributed  filter, consensus, sensor networks, performance analysis, directed graph.
\end{IEEEkeywords}

\section{Introduction}

In the preceding two decades,  investigations into sensor networks have occupied a prominent position  in the realm of systems and control, owing to its extensive applications  encompassing health care monitoring \cite{lin2020wireless,abdulkarem2020wireless},   environmental sensing \cite{corke2010environmental,hart2006environmental}, and collaborative   mapping \cite{gao2020random,pei2020decorrelated}.
In order to meet the demands for the network reliability and alleviate  the computational and communicative burdens on energy-constrained sensors, the implementation of distributed state estimation serves as a good solution to these challenges.
 The objective of distributed state estimation is to estimate the state of the target system for each sensor in sensor networks.

Building on  various dynamical systems and constraint conditions,  a wealth of  research   has  emerged  in the field of  distributed state estimation.
The evolution of consensus theory within the domain of  multi-agent systems
has imparted   profound insights  into the information fusion  for distributed state estimation
\cite{olfati2004consensus,li2009consensus,wen2015containment}.
Several reputable consensus-based approaches to distributed state estimation have been  proposed
\cite{olfati2007distributed,yang2017stochastic,battilotti2020asymptotically,battistelli2014kullback,
olfati2009kalman,battistelli2014consensus,duan2022distributed}.
Depending on  the    filter structures,  these filters  fuse    different types of  information    from local neighbors, including local state estimates \cite{battistelli2014kullback,yang2017stochastic,battilotti2020asymptotically},    intermediate  quantities \cite{olfati2009kalman,battistelli2014consensus}, or  measurements \cite{duan2022distributed}.
For the fusion of   local state estimates,
Battistelli et al.~\cite{battistelli2014kullback}   introduced   a  distributed filter based  on  the   covariance intersection method, which combines estimates  from different sensors. In another approach,
Battilotti et al. \cite{battilotti2020asymptotically}   developed    a  distributed filter  that fuses   the   state estimation deviations among  neighboring sensors.
Regarding the  fusion  of  intermediate  quantities,
Olfati-Saber~\cite{olfati2009kalman}    designed  a
Kalman-consensus filter using  intermediate quantities  to  approximate  the centralized filter,  with a Lyapunov-based stability analysis.
Additionally,  Battistelli et al.  \cite{battistelli2014consensus} presented a distributed filter  that  fuses   both state estimates  and  intermediate quantities, thereby   preserving   the  benefits   of both approaches.
For the  fusion of   local  measurements,  Duan et al.~\cite{duan2022distributed}    proposed
 a distributed state estimation filter    for continuous-time systems with correlated measurement noise.

However,  compared to   existing literature,   the research on   the  fusion of local measurements has not been  thoroughly  investigated   for discrete-time linear  systems due to the following challenges:
\begin{enumerate}
\item   Designing   the  filter structure   and  information fusion strategies   using local measurements  is    challenging    while   ensuring  the filter performance.

\item  Existing  gain  design  methods  for   directed graphs    \cite{battistelli2014kullback,battistelli2014consensus}  face complexities  in applications,   and developing an effective distributed design  method is crucial.

\item  It is necessary to explore  the performance gap between the proposed filter and the centralized filter, with a particular  focus on  the convergence rate  regarding   the fusion step number.
\end{enumerate}

Motivated by the above  observations,  this paper aims to propose a  novel  consensus-based   distributed filter over directed graphs    using a dynamic averaging algorithm,    and provides  a   comprehensive performance analysis of  the proposed distributed  filter.
The main contributions of this paper are summarized below:
\begin{enumerate}
\item  We propose a  novel   consensus-based     distributed     filter   using   an augmented  leader-following information fusion strategy  over directed graphs  under the collectively observability condition.
    Compared to the existing literature \cite{olfati2007distributed,yang2017stochastic,battilotti2020asymptotically,battistelli2014kullback,
olfati2009kalman,battistelli2014consensus,duan2022distributed},   this filter  provides a novel filter structure that has  been  rarely studied.

\item  Under the coupling between  the measurement estimation error  and  the state estimation error,
 we provide   gain design methods that  rely   solely   on    local  information. Additionally,  we  derive   the lower bound for  the fusion step number  (\textbf{Theorem~\ref{theo l0 l}}).

\item   We analyze  the  impact of  the fusion step number  on the steady-state performance.
     Specifically,  we provide
    the lower  bounds  on  the convergence rates  of the  steady-state performance  gap between the proposed  distributed filter  and the centralized filter as  the fusion step approaches  infinity (\textbf{Theorem~\ref{theorem exponential convergence}}).
\end{enumerate}

The remainder of this paper is organized as follows.  Section~\ref{sec pre and problem fomulation} presents  the  preliminaries and  problem formulation.
Section~\ref{sec distributed filter} proposes  a  consensus-based  distributed
filter over directed graphs, and  provides  two parameter design methods.
Section~\ref{sec performance analysis} analyzes  the steady-state performance of the proposed filter  as  the  fusion step number increases.
Section~\ref{sec simulations} presents  two numerical  examples   to validate the   effectiveness of the results.  Finally,  Section~\ref{sec conclusion}  concludes the paper.

\textit{Notations}:
For a matrix $A\in \mathcal{R}^{n\times n}$,
 $\Vert A\Vert_2$ is the spectral norm,  $\rho(A)$ is the spectral radius, and $[A]_{ij}$ denotes the $(i,j)$-th element of the matrix $A$.
   Symbol  $\otimes$  represents the Kronecker product.
 The matrix inequalities $A > B$  and  $A\geq B$ signify that $A-B$ is positive definite and positive semi-definite, respectively.

\section{Preliminaries and Problem Statement}\label{sec pre and problem fomulation}

\subsection{Some Useful Lemmas}

\begin{definition}\cite{horn2012matrix}\label{definition irreducibly diagnoally dominant}
For a matrix $M\in \mathcal{R}^{n\times n}$, $M$ is said to be irreducibly diagonally dominant if 1)  $M$ is irreducible, 2) 
$M$ is diagonally dominant, i.e., $\vert m_{ii} \vert\geq R^{s}_i(M)$ for all $i = 1,\ldots, n$, 3) There is an $i\in \{1,\ldots,n\}$ such that $\vert m_{ii} \vert> R^{s}_i(M) $,
where $R^{s}_i(M) = \sum_{j\neq i}\vert m_{ij}\vert$.
\end{definition}

\begin{lemma}\cite{horn2012matrix}\label{lemma domint matrix property}
Let the matrix  $M$ be irreducibly diagonally dominant. Then,
1) $M$ is nonsingular,   2)   If $M$ is Hermitian and every main diagonal entry is positive,  $M$ is positive definite.
\end{lemma}

\begin{lemma}\cite{horn2012matrix}\label{lemma Akto0}
For a  matrix $M$ and a positive integer $k$, it holds $\lim_{k\to \infty}M^k =0$ if and only if $\rho(M)<1$.
\end{lemma}
\begin{lemma}\cite{horn2012matrix}\label{lemma rho A by sum element}
For a nonnegative matrix $M=[m_{ij}]$, it holds   $\text{min}_{1\leq i\leq n}\sum^n_{j=1}m_{ij}\leq \rho(M)\leq \text{max}_{1\leq i\leq n}\sum^n_{j=1}m_{ij}.$
\end{lemma}

\begin{lemma}\cite{qian2022consensus}\label{lemma norm of Ak}
For any matrix $M\in \mathcal{R}^{n\times n}$, it holds
$\Vert M^k\Vert_2 \leq \sqrt{n}\sum^{n-1}_{j=0}\binom{n-1}{j}\binom{k}{j}\Vert M\Vert^j_2 \rho(M)^{k-j},$
where $\binom{m}{n}$ is the combinatorial number with $\binom{k}{j}=0$ for $j>k$.
\end{lemma}

\subsection{System Model}
Consider a discrete-time linear time-invariant system observed by a network of $N$ sensors:
\begin{equation}\label{eq dynamics}
\begin{aligned}
& x_{k+1} = Ax_{k}+\omega_{k},\\
& y_{i,k} = C_{i}x_{k}+\nu_{i,k}, ~~~i=1,2,...,N,
\end{aligned}
\end{equation}
where  $k$  represents    the discrete-time index, $i$  represents  the $i$-th sensor  in  the network,  $x_k\in \mathcal{R}^n$ is the system state vector, $y_{i,k}\in \mathcal{R}^{r_i}$ is the measurement vector of  sensor $i$,  $A\in \mathcal{R}^{n\times n}$ is the state transition matrix, $C_i\in \mathcal{R}^{r_i\times n}$ is the observation matrix,  and  $\omega_k\in \mathcal{R}^{n}$ and $\nu_{i,k}\in \mathcal{R}^{r_i}$ are  zero-mean Gaussian noise with the covariances $Q_k\in \mathcal{R}^{n\times n}$ and $R_{i,k}\in\mathcal{R}^{r_i\times r_i}$, respectively.
The noise sequences $\{\omega_k\}^{\infty}_{k=0}$ and $\{\nu_{i,k}\}^{\infty,N}_{k=0,i=1}$ are mutually uncorrelated.
 Let $C=[C^T_1,\ldots,C^T_N]^T \in \mathcal{R}^{r\times n}$  denote  the augmented  observation matrix, and
$R=\text{diag}\{R_1,\ldots,R_N\}\in \mathcal{R}^{r\times r}$  represent    the  augmented  measurement noise covariance matrix, where   $r=\sum^N_{i=1}r_i$. The communication topology of  the sensor network is  described by a graph   ${\mathcal{G}}(\mathcal{V},\mathcal{E})$, where   $\mathcal{V}=\{1,2,\ldots,N\}$  denotes  the sensor set, and
  $\mathcal{E} \subseteq \mathcal{V} \times \mathcal{V}$  represents  the edge set  defining  the communication links among the sensors.
For $i, j \in \mathcal{V}$,  the edge  $(j,i)$   indicates   that sensor  $j$ can transmit information to sensor  $i$, and sensor  $j$ is  referred as  a neighbor of sensor  $i$. The  neighbor set  of sensor  $i$ is denoted  as  $\mathcal{N}_i =\{j|(j,i)\in \mathcal{V}\}$,  and  $\vert\mathcal{N}_i\vert$  represents the number of neighbors of sensor~$i$.
The adjacency matrix is   defined as  $S = [a_{ij}]_{N\times N}$, where
 $a_{ij} = 1$ if $(j,i)\in \mathcal{E}$ and   $a_{ij} = 0$ otherwise.
 Let $D = \text{diag}\{\vert\mathcal{N}_1\vert,\ldots,\vert\mathcal{N}_N\vert \}$,   and  the Laplacian matrix  is  defined as   $\mathcal{L} = D-S$.
  An edge $(i,j)$ is  undirected if $(i,j)\in \mathcal{E}$  implies    $(j,i) \in \mathcal{E}$. The  communication graph is termed  undirected if  all edges  are  undirected.   The graph $G$ contains  a directed path  from sensor~$i_1$ to sensor~$i_m$, if there exists a sequence of connected edges $(i_k, i_{k+1}), k=1,\ldots,m-1$.
 The communication graph is called strongly connected, if there exists a path between any pair of distinct nodes.

\begin{assumption}\label{ass observable}
$(C, A)$ is observable.
\end{assumption}

\begin{assumption}\label{ass graph connected}
The communication graph is directed and  strongly  connected.
\end{assumption}

\begin{remark}
Assumption~\ref{ass observable} shows that    observability can be achieved  by  the entire network, even if it is not satisfied
by  any  single sensor.
Assumption~\ref{ass graph connected}
can be extended to a  jointly connected switching  topology  by applying   the corresponding definition of communication networks \cite{su2012two,liu2020discrete}.
\end{remark}

\subsection{Problem Statement}
\begin{enumerate}
\item   Develop   a  distributed  filter  over directed graphs with a leader-following information fusion strategy using local measurements.  Explore   parameter design methods to guarantee the filter's performance.
\item    Find out the performance  gap  between the proposed distributed filter and the centralized filter. Evaluate how the fusion step  number  influences  the performance.
\end{enumerate}

\section{Distributed Filter}\label{sec distributed filter}

In this section,   we propose  a novel consensus-based distributed filter.  Subsequently, we define the state  estimation  error and the measurement estimation error, and  analyze  their properties.  Furthermore,  we provide  two parameter design methods  to guarantee the  filter's performance.

\subsection{Design of  the Distributed Filter}

This subsection proposes a novel distributed state estimator  over directed graphs  with  the consensus-based information fusion  strategy.

First,  design  sensor $i$'s estimate of  sensor $j$'s measurement at the $l$-th fusion step, denoted as $z^{(l)}_{ij,k}$,   as follows: 
 \begin{equation}\label{eq estimate measurement zero single}
\begin{aligned}
z^{(0)}_{ij,k}=C_jA \hat x_{i,k-1|k-1},~~~~~~~~~~~~~~~~~~~~~~~~~~~
\end{aligned}
\end{equation}
\begin{equation}\label{eq estimate measurement lstep single}
\begin{aligned}
z^{(l)}_{ij,k}& = z^{(l-1)}_{ij,k}-\mu_{ij}\Big[\sum_{t\in \mathcal{N}_i }a_{it}(z^{(l-1)}_{ij,k}-z^{(l-1)}_{tj,k})\\
&~~~~+a_{ij}(z^{(l-1)}_{ij,k}-y_{j,k})\Big], ~~~l=1, 2,\ldots,
\end{aligned}
\end{equation}
where  the consensus gain  $\mu_{ij}$,  to be designed  later, is a positive constant, and $a_{ij}$ is the element of the adjacency matrix corresponding to the  communication topology.
By denoting the augmented vectors as
$z^{(l)}_{i,k} = [(z^{(l)}_{i1,k})^T,\ldots,(z^{(l)}_{iN,k})^T]^T$ and
$y_k = [y^T_{1,k},\ldots,y^T_{N,k}]^T$,  $z^{(l)}_{i,k}$ can be expressed as
 \begin{equation}\label{eq estimate measurement zero augmented}
\begin{aligned}
z^{(0)}_{i,k}=CA \hat x_{i,k-1|k-1},~~~~~~~~~~~~~~~~~~~~~~~~~~~~~
\end{aligned}
\end{equation}
\begin{equation}\label{eq estimate measurement lsteo augmented}
\begin{aligned}
~~~~z^{(l)}_{i,k}& = z^{(l-1)}_{i,k}-\Lambda_i \Big[\sum^N_{t=1}a_{it}(z^{(l-1)}_{i,k}-z^{(l-1)}_{t,k})~~~~\\
&~~~~+ B_i(z^{(l-1)}_{i,k}-y_{k})\Big],
\end{aligned}
\end{equation}
where $\Lambda_i =\text{diag}\{\mu_{i1}\otimes I_{r_1},\ldots,\mu_{iN}\otimes I_{r_N}\}$ and      $B_{i} = \text{diag}\{a_{i1}\otimes I_{r_1},\ldots,a_{iN}\otimes I_{r_N} \}$.

Next, the state estimator structure  of the  system  (\ref{eq dynamics}) for  sensor $i$ is designed as
\begin{equation}\label{eq filter update}
\begin{aligned}
\hat x_{i,k|k} = (A-KCA)\hat x_{i,k-1|k-1} + Kz^{(l)}_{i,k},~~~~~
\end{aligned}
\end{equation}
where  the gain  matrix $K$ is  given by   $K=\bar P C^TR^{-1},$
$\bar P = P-PC^T(CPC^T+R)^{-1}CP$,  and  $P$ is determined by solving the discrete algebraic Riccati equation $P = APA^T+Q-APC^T(CPC^T+R)^{-1}CPA^T$.

\begin{figure}[!htb]
	\centering
	{\includegraphics[width=2.6in]{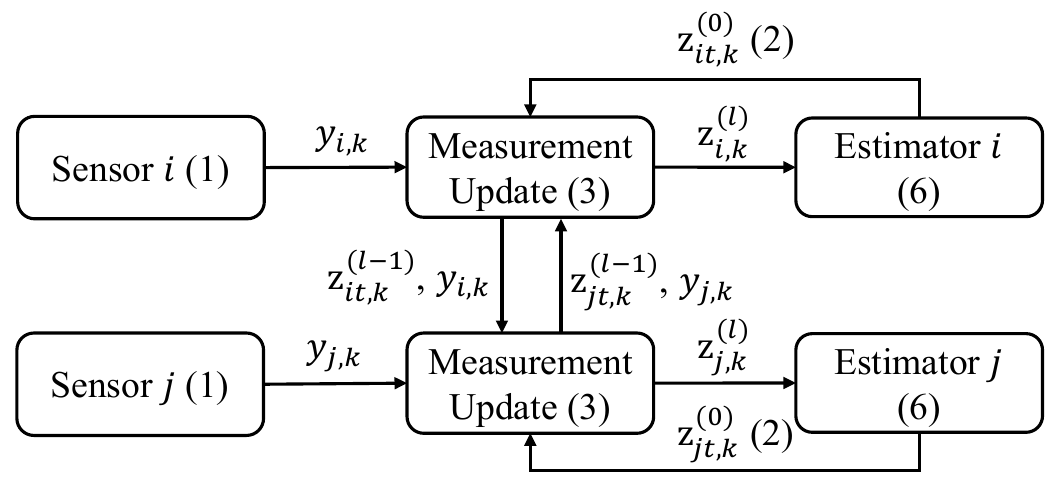}}
	\caption{Illustration  of the proposed distributed estimator.}
	\label{fig_algorithm_illustration}
\end{figure}

Now, the distributed  estimator
 is constructed   based on    (\ref{eq filter update}), and its structure   is illustrated  in Fig.~\ref{fig_algorithm_illustration}.

\begin{remark}
In the algorithm, both  the measurement estimates $z^{(l-1)}_{it,k}$ and the raw  measurements $y_{i,k}$
are transmitted. 
Between  two   consecutive    time steps,   the information is exchanged   $l$ times.   Since  each node  may not obtain  sufficient  measurements  from its neighbors, 
  the measurement estimates,  obtained  through prediction, compensate for this limitation.
\end{remark}

\begin{remark}
	The proposed distributed  estimator  offers several  advantages. First, it  complements   existing  distributed  filters  through  a local  measurement prediction fusion mechanism,  thereby establishing  a general framework  for further  investigation. 
	Second, the proposed distributed filter demonstrates strong performance  by  achieving   asymptotically   optimality    as the fusion step $l$ tends to infinity.
	Third,  it  can be  applied to  directed graphs with a simpler  parameter design method compared to \cite{battistelli2014kullback,battistelli2014consensus}. 
	Fourth, when direct measurement information
	is unavailable, the predicted measurements  compensate for missing data, thus
	maintaining estimation performance.
	Finally, its  flexibility in communication steps  makes it suitable for  extensions to event-triggered or adaptive communication strategies, while the consensus framework can also be extended to support privacy-preserving and secure filtering schemes.
\end{remark}

\subsection{Two Estimation Errors}

This subsection computes  two estimation  errors for further analysis.
Define  the measurement estimation error and the state  estimation error  as  $\varepsilon^{(l)}_{i,k} = z^{(l)}_{i,k}-y_{k},$ and 
 $e_{i,k|k}=\hat x_{i,k|k}-x_k,$
respectively. 
Denote  the augmented  vectors as
\begin{equation*}
\begin{aligned}
&e_{k} = [e^T_{1,k}, \ldots, e^T_{N,k}]^T, ~~~~~~~\varepsilon^{(l)}_{k} = [(\varepsilon^{(l)}_{1,k})^T, \ldots, (\varepsilon^{(l)}_{N,k})^T]^T,\\
&z^{(l)}_{k} = [(z^{(l)}_{1,k})^T,\ldots,(z^{(l)}_{N,k})^T]^T, ~ \nu_k=[\nu^T_{1,k},\ldots,\nu^T_{N,k}]^T.
\end{aligned}
\end{equation*}
Next, we derive the  statistical  properties   of   the  measurement estimation error $\varepsilon^{(l)}_{k}$  and  the state estimation error  $e_{i,k|k}$.  Additionally, we define their estimation error covariances as  $P_{\varepsilon,k}= E\{\varepsilon^{(l)}_{k}(\varepsilon^{(l)}_{k})^T\}$ and $P_{k|k} = E\{e_ke^T_k\}$.

\begin{proposition}\label{lemma measurement estimation error}
 The following statements hold
\begin{enumerate}
\item The measurement estimation error   $\varepsilon^{(l)}_{k}$ is
\begin{equation*}
\begin{aligned}
\varepsilon^{(l)}_{k} = G^{l}\varepsilon^{(0)}_{k},~~~~~~~~~~~~~~~~~~~~~~~~
\end{aligned}
\end{equation*}
where $G = I_{Nr} -\Lambda(\mathcal{L}\otimes I_r +B)$,   $G^l$	represents  the 
	$l$-th power of $G$,   $\Lambda=\text{diag}\{\Lambda_1,\ldots,\Lambda_N\}$, $B = \text{diag}\{B_1,\ldots,B_N\}$, and
\begin{equation*}
\begin{aligned}
\varepsilon^{(0)}_{k}
&= (I_N\otimes CA)
e_{k-1}-1_N\otimes\nu_k\\
&~~~~-(I_N\otimes C)(1_N\otimes\omega_{k-1}).
\end{aligned}
\end{equation*}

\item The measurement estimation error covariance $P_{\varepsilon,k}$ is
\begin{equation}\label{eq lemma epsilon P}
\begin{aligned}
P_{\varepsilon,k}
&=G^l(I_N\otimes CA)P_{k-1}(I_N\otimes CA)^T(G^l)^T\\
&~~~~+  G^l(I_N\otimes C)(U_N\otimes Q)(I_N\otimes C)^T(G^l)^T~~\\
&~~~~+ G^l(U_N\otimes R)(G^l)^T.
\end{aligned}
\end{equation}

\item    The state estimation error $e_k$ is
\begin{equation*}
\begin{aligned}
~~e_{k} = \mathcal{A}(l) e_{k-1}-\mathcal{B}(l)(1_N\otimes \omega_{k-1})+\mathcal{D}(l)(1_N\otimes \nu_k),
\end{aligned}
\end{equation*}
where
\begin{equation}\label{eq lemma state estimation error A}
\begin{aligned}
~\mathcal{A}(l)=I_N \otimes(A-KCA)+(I_N \otimes K)G^l(I_N\otimes CA),
\end{aligned}
\end{equation}
\begin{equation}\label{eq lemma state estimation error B}
\begin{aligned}
\mathcal{B}(l)=I_N\otimes(I-KC)-(I_N \otimes K)G^l(I_N\otimes C),~~~
\end{aligned}
\end{equation}
and
\begin{equation}\label{eq lemma state estimation error D}
\begin{aligned}
\mathcal{D}(l)=(I_N \otimes K)(I_{Nr}-G^l).~~~~~
\end{aligned}
\end{equation}

\item  The state estimation error covariance $P_{k|k}$  is
\begin{equation}\label{eq lemma state estimation error P}
\begin{aligned}
~~P_{k|k}= \mathcal{A}(l)P_{k-1|k-1}\mathcal{A}^T(l)+ \Phi(l),
\end{aligned}
\end{equation}
where
\begin{equation}\label{eq lemma state estimation error phi}
\begin{aligned}
\Phi(l) = \mathcal{B}(l)(U_N\otimes Q)\mathcal{B}^T(l)+\mathcal{D}(l) (U_N\otimes R)  \mathcal{D}^T(l).
\end{aligned}
\end{equation}
\end{enumerate}
\end{proposition}

The proof of Proposition \ref{lemma measurement estimation error} is given in  Appendix \ref{app lemma measurement estimation error}.

\begin{remark}
Proposition~\ref{lemma measurement estimation error} shows that
$\varepsilon^{(l)}_k$ is a random variable with  the mean  $E\{\varepsilon^{(l)}_{k}\}= G^{l}(I_N\otimes CA)E\{e_{k-1}\}$  and  the covariance $P_{\varepsilon,k}$.   Under the assumption that $\rho(G)<1$, as $l$ tends to infinity,  it holds $\lim_{l\to \infty} E\{\varepsilon^{(l)}_{k}\} =0$ and  $\lim_{l\to \infty} P_{\varepsilon,k} =0$.
The state  estimation error $e_k$ and the measurement estimation error $\varepsilon^{(l)}_k$  interact and influence each other.  To improve the filter performance,  two parameters  will  be designed:  one is the parameter $\mu_{ij}$ in  $G$,  and the other  is the fusion step $l$.
\end{remark}

\subsection{Parameter Design for $\mu_i$}\label{subsec parameter mu}
This subsection will  design  the parameter $\mu_i$   to ensure the convergence of two estimation errors.

\subsubsection{Distributed  Design}
We propose  a distributed design approach for $\mu_{ij}$  to circumvent the utilization of the global topology  information for each sensor.
First, define the submatrix $G_{[ij]}$ as
$G_{[ij]} = [0_{r\times r(i-1)}, I_{r}, 0_{r\times r(N-i)}] G  [0^T_{r\times r(j-1)}, I^T_{r}, 0^T_{r\times r(N-j)}]^T,$
where $[ij]$ represents the $i$-th row and $j$-th column index of the  corresponding matrix block. The detailed submatrices are
$G_{[ii]} = \text{diag}\{g_{ii,1},\dots, g_{ii,N} \}$  and $G_{[ij]} =   \text{diag}\{g_{ij,1},\dots, g_{ij,N} \}$, where
\begin{equation}\label{eq giil}
\begin{aligned}
~~~~g_{[ii,l]} &= (1-\mu_{il}(l_{ii}+a_{il}))\otimes I_{r_l},
\end{aligned}
\end{equation}
\begin{equation}\label{eq gijl}
\begin{aligned}
g_{[ij,l]} &= (-\mu_{il}l_{ij})\otimes I_{r_l},~~~~~~~
\end{aligned}
\end{equation}
and $l_{ij}$ is the $i$-th row and $j$-th column element of the Laplacian   matrix $\mathcal{L}$.
Next, we   introduce   a method  to  select  the parameter  $\mu_{ij}$    from  a reasonable range, given by:
\begin{equation}\label{eq uil range}
0<\mu_{il}\leq \frac{1}{l_{ii}+a_{il}}.~~~~~~~~~~~~
\end{equation}

We provide the following lemma to illustrate  the choice of    $\mu_{il}$ satisfying  (\ref{eq uil range}) can ensure   $\rho(G)<1$.

\begin{lemma}\label{lemma mu range}
Under Assumption \ref{ass graph connected}, if $\mu_{il}$ satisfies (\ref{eq uil range})
then it holds   $\rho(G)<1$.
\end{lemma}

The proof of Lemma \ref{lemma mu range} is given in  Appendix \ref{app lemma mu range}.

\vspace{6pt}
\subsubsection{Unified Design}
A unified design method is also provided for undirected graphs.
Consider the scenario where all $\mu_i$ are identical, denoted as $\mu$.  The symbol $G$ is then redefined   as   $G = I_{Nr} -\mu(\mathcal{L}\otimes I_r +B).$
Based on Lemma~\ref{lemma domint matrix property}, $G$ is positive definite.
 By selecting    $\mu < \frac{1}{\rho(\mathcal{L}\otimes I_r +B)}$,  it is ensured that $\rho(G)<1$.
 This method  requires   the global communication  topology  information.

\subsection{Parameter Design for $l$}

This subsection aims to  provide  the lower bound  of   the  fusion step number  $l$
to ensure the convergence of  the distributed estimators.

\begin{theorem}\label{theo l0 l}
Under Assumptions~\ref{ass observable} and  \ref{ass graph connected},  if $\Vert G \Vert_2<1$ and $l> l_0$,
where
\begin{equation}
\begin{aligned}
~l_0=\text{log}_{\Vert G\Vert_2}\frac{1-\Vert A-KCA\Vert_2}{\Vert K\Vert_2\Vert CA\Vert_2},
\end{aligned}
\end{equation}
then  $\mathcal{A}(l)$ is Schur stable, and  
the estimation error covariance $P_{i,k|k}$  is uniformly upper-bounded.

\end{theorem}

The proof of Theorem \ref{theo l0 l} is given in  Appendix \ref{app theo l0 l}.

\begin{remark}
 From Theorem \ref{theo l0 l},  the value $l_0$ is determined by two factors:
 $\Vert G\Vert_2$ and $\bar K = \frac{1-\Vert A-KCA\Vert_2}{\Vert K\Vert_2\Vert CA\Vert_2}$.
  On  one hand, the term  $\Vert G\Vert_2$ is related  to  the communication graph and the parameter $\mu_i$.  The smaller  $\Vert G\Vert_2$ is , the smaller $l_0$  becomes. On the other hand, the term  $\bar K$  indicates  the stability margin of the estimator.  The larger  $\bar K$ is, the smaller $l_0$ becomes.
\end{remark}

\section{Performance Analysis}\label{sec performance analysis}
This section    analyzes  the  performance of the proposed distributed filter. Specifically,   we present
the steady-state performance gap between the centralized filter and the proposed distributed filter as  the  fusion step number $l$ increases.

\subsection{Centralized Estimator}\label{sec centrailized estimator}
Consider a centralized estimator  for  comparison, given by:
\begin{equation}\label{eq centrailized estimaitor update}
\begin{aligned}
\hat x_{c,k|k} = A\hat x_{c,k-1|k-1} + K( y_k- CA\hat x_{c,k-1|k-1}),
\end{aligned}
\end{equation}
Denote the estimation error as $e_{c,k|k}=\hat x_{c,k|k}-x_k$,
the  estimation error covariance as
$P_{c,k|k}=E\{e_{c,k|k}e^T_{c,k|k}\}$,
the augmented estimation error as $e_{cc,k} = 1_N\otimes e_{c,k}$,
and  the augmented  estimation error covariance as
$P_{cc,k|k}=E\{e_{cc,k|k}e^T_{cc,k|k}\} = U_N\otimes P_{c,k|k}.$
By combining the estimator  (\ref{eq centrailized estimaitor update}) and the dynamical system~(\ref{eq dynamics}),
$e_{cc,k} = 1_N\otimes e_{c,k}$  can be computed as
\begin{equation}\label{eq lemma centrailzed error}
\begin{aligned}
~~~e_{cc,k} = \mathcal{A}_{cc} e_{cc,k-1}-\mathcal{B}_{cc}(1_N\otimes \omega_{k-1})+\mathcal{D}_{cc}(1_N\otimes \nu_k),
\end{aligned}
\end{equation}
where
\begin{equation}\label{eq Acc}
\begin{aligned}
\mathcal{A}_{cc}=I_N \otimes(A-KCA),
\end{aligned}
\end{equation}
\begin{equation}\label{eq Bcc}
\begin{aligned}
\mathcal{B}_{cc}=I_N\otimes(I-KC),~~
\end{aligned}
\end{equation}
and
\begin{equation}\label{eq Dcc}
\begin{aligned}
\mathcal{D}_{cc}=I_N \otimes K.~~~~~~~~~~~~
\end{aligned}
\end{equation}
The corresponding estimation error covariance $P_{cc,k|k}$  is
\begin{equation}\label{eq lemma centralized error covariance}
\begin{aligned}
P_{cc,k|k}
& = \mathcal{A}_{cc}P_{cc,k-1|k-1}\mathcal{A}^T_{cc}+ \Phi_{cc},~~~~~~~~~~
\end{aligned}
\end{equation}
where
\begin{equation}\label{eq phicc}
\begin{aligned}
\Phi_{cc} = \mathcal{B}_{cc}(U_N\otimes Q)\mathcal{B}^T_{cc}+\mathcal{D}_{cc} (U_N\otimes R)  \mathcal{D}^T_{cc}.~~
\end{aligned}
\end{equation}
Since we analyze the  steady-state performance  gap,    the centralized filter with  the fixed gain
will converge to  the optimal  solution.

\subsection{Convergence Analysis}

This section   investigates   the  steady-state  performance gap  between $P_{cc,k|k}$ and $P_{k|k}$,  and sheds  light on  the influence  of  the fusion step number  $l$ on the  performance.
Section~\ref{subsec parameter mu} has shown that the parameter $\mu_{ij}$  can be designed  to ensure that  $\rho(G)<1$. Therefore, we     make the following assumption.

\begin{assumption}\label{ass G lees than 1}
The  matrix $G$ satisfies $\rho(G)<1$, and $\mathcal{A}(l)$ is  Schur stable.
\end{assumption}

\begin{remark}
Assumption~\ref{ass G lees than 1}  can be satisfied 
through   the proposed parameter design method,  as  established  in Lemma~\ref{lemma mu range} and Theorem~\ref{theo l0 l}.  In other words, this assumption  indicates that the  parameters have been appropriately designed.
\end{remark}

\begin{theorem}\label{theo converge to DLE}
Under Assumptions \ref{ass observable}, \ref{ass graph connected},  and \ref{ass G lees than 1},
$P_{k|k}$  and $P_{cc,k|k}$   converge to  the unique solutions of the discrete-time Lyapunov equations (DLE)
\begin{equation}
\begin{aligned}
P^{(l)} & = \mathcal{A}(l)P^{(l)}\mathcal{A}^T(l)+ \Phi(l),~~
\end{aligned}
\end{equation}
and
\begin{equation}
\begin{aligned}
P_{cc}
& = \mathcal{A}_{cc}P_{cc}\mathcal{A}^T_{cc}+ \Phi_{cc},~~~~~~~~
\end{aligned}
\end{equation}
respectively, where $\mathcal{A}(l)$,   $\Phi(l)$,  $\mathcal{A}_{cc}$, and $\Phi_{cc}$    have been defined in (\ref{eq lemma state estimation error A}), (\ref{eq lemma state estimation error phi}), (\ref{eq Acc}), and  (\ref{eq phicc}), respectively.
\end{theorem}

The proof of Theorem \ref{theo converge to DLE} is given in  Appendix \ref{app theo converge to DLE}.

To evaluate the  effect of the  fusion step number  $l$ on the steady-state performance,  we introduce  the following notations
\begin{equation}
\begin{aligned}
& \bar {\mathcal{A}}(l) = \mathcal{A}(l)-\mathcal{A}_{cc}= (I_N \otimes K)G^l(I_N\otimes CA),\\
& \mathcal{\bar B}(l) = \mathcal{B}(l)-\mathcal{B}_{cc}= -(I_N \otimes K)G^l(I_N\otimes C),\\
& \mathcal{\bar D}(l) = \mathcal{D}(l)-\mathcal{D}_{cc}= -(I_N \otimes K)G^l.
\end{aligned}
\end{equation}

\begin{lemma}\label{lemma differnce series}
The difference between $P^{(l)}$ and $P_{cc}$  is
\begin{equation}\label{eq lemma pl-pcc formal}
\begin{aligned}
 P^{(l)}  - P_{cc} = \sum^{\infty}_{k=0} \mathcal{A}^{k}_{cc} \bar \Phi(l) (\mathcal{A}^T_{cc})^k,~~
\end{aligned}
\end{equation}
where
\begin{equation}\label{eq lemma bar phi fin}
\begin{aligned}
 \bar \Phi(l) & = {\mathcal{\bar A}}(l)P^{(l)}\mathcal{A}^T_{cc}+\mathcal{A}_{cc}P^{(l)} {\mathcal{\bar A}}^T(l)+{\mathcal{\bar A}}(l)P^{(l)}{\mathcal{\bar A}}^T(l)\\
 &~~~+{\mathcal{\bar B}}(l)(U_N\otimes Q)\mathcal{B}^T_{cc}+\mathcal{B}_{cc}(U_N\otimes Q) {\mathcal{\bar B}}^T(l)\\
 &~~~+{\mathcal{\bar B}}(l)(U_N\otimes Q){\mathcal{\bar B}}^T(l)+{\mathcal{\bar D}}(l)(U_N\otimes R)\mathcal{D}^T_{cc}\\
 &~~~+\mathcal{D}_{cc}(U_N\otimes R) {\mathcal{\bar D}}^T(l)+{\mathcal{\bar D}}(l)(U_N\otimes R){\mathcal{\bar D}}^T(l).
\end{aligned}
\end{equation}
\end{lemma}

The proof of Lemma \ref{lemma differnce series} is given in  Appendix \ref{app lemma differnce series}.

\begin{theorem}\label{theorem exponential convergence}
Under Assumptions    \ref{ass observable}  and   \ref{ass graph connected},        there exist constants $M_1, M_2>0$, such that
\begin{enumerate}
\item  If $\rho(G)<1$, it holds
\begin{equation}\label{eq rhoG 1 norm}
\begin{aligned}
\Vert P^{(l)}  - P_{cc}\Vert_2 \leq M_1 l^{Nr}\rho(G)^{l-Nr}.~~~~
\end{aligned}
\end{equation}

\item  If  $\Vert G \Vert_2<1$, it holds
\begin{equation}\label{eq rhoG 1 norm 2}
\begin{aligned}
\Vert P^{(l)}  - P_{cc}\Vert_2 \leq M_2 \Vert G\Vert^l_2.~~~~~~~~~~~~~
\end{aligned}
\end{equation}

\end{enumerate}

\end{theorem}

The proof of Theorem \ref{theorem exponential convergence} is given in  Appendix \ref{app theorem exponential convergence}.

\begin{remark}
Two  conditions related to the communication  graph  are presented,
since
ensuring that the spectral radius of the matrix is less than 1 does not necessarily imply that its spectral norm is also less than 1.
For an undirected graph,   the spectral radius  $\rho(G)$  equals  the spectral norm  $\Vert G \Vert_2$, and  the conditions  in  item (1) and (2) of Theorem \ref{theorem exponential convergence} are identical.
In both cases,
as the fusion step number  $l$   approaches  infinity, $\Vert P^{(l)}  - P_{cc}\Vert_2$ converges to $0$, as demonstrated in  (\ref{eq rhoG 1 norm})~and~(\ref{eq rhoG 1 norm 2}).

\end{remark}

\section{Simulations}\label{sec simulations}

\begin{figure}[!htb]
\centering
{\includegraphics[width=1.2in]{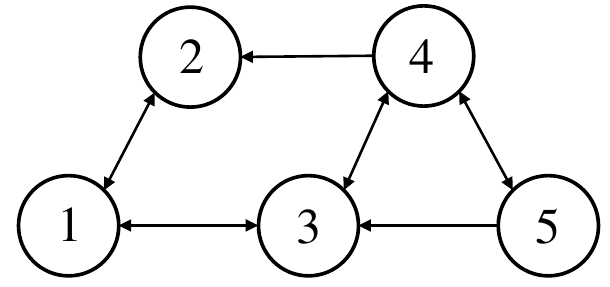}}
\caption{The diagram of the communication topology.}
\label{fig communication_topology simulation 1}
\end{figure}

In this section,  the effectiveness of the theoretical results  is validated  through a target tracking numerical experiment. Consider
a sensor network comprising five sensors labeled from 1 to 5, and its communication topology is illustrated  in Fig. \ref{fig communication_topology simulation 1}.

Consider the target dynamics described by   $A= \text{diag}\{I_2~ TI_2; 0~ I_2\}$,
where $T=0.25$ is the discretization interval. The process noise covariance is  defined as
\begin{equation}
\begin{aligned}
\bar Q = \left (\begin{array}{cc}
\frac{T^3}{3} & \frac{T^2}{2} \\
\frac{T^2}{2} & T \end{array}\right),~~~ Q = \left (\begin{array}{cc}
\bar Q & 0.5 \bar Q \\
0.5 \bar Q & \bar Q \end{array}\right).
\end{aligned}
\end{equation}
Two kinds of sensors are employed in the sensor network:  position  sensors and velocity sensors.  The observation matrix for the
 position sensors  is given by  $C_p = [I_2~0_2]$
with  the measurement noise covariance $R_p = \text{diag}\{1, 1 \}$. Similarly, the observation matrix for the velocity sensors is represented by
$C_v = [0_2~I_2]$
with the measurement noise covariance $R_v = \text{diag}\{5, 5\}$.
In this sensor network,  sensors 1, 2, and 4 are designated  as  the  position sensors, while  sensors 3 and 5 serve as  the  velocity sensors.
The initial state is set as  $x_{0}= [1;1;1;1]$, and  the initial state estimate  is   set as a random variable  with  the mean $\hat x_{i,k|k} = x_{0} $  and the initial estimation covariance $P_{i,0|0} = \text{diag}\{10,10,10,10\}$.   The parameter  $\mu_{ij}$ is selected as   $\frac{1}{l_{ii}+a_{il}+1}$.
The mean square  error (MSE)   is utilized to evaluate  the performance of the estimator based on the Monte Carlo method, described by
\begin{equation}
\begin{aligned}
MSE_{i,k} = \frac{1}{M} \sum^{M}_{m=1}\Vert \hat x^{(m)}_{i,k} - x^{(m)}_{k} \Vert^2_2,
\end{aligned}
\end{equation}
where $M$ is the trial number, and $\hat x^{(m)}_{i,k}$ and $x^{(m)}_{k}$
are the state estimate and the true state at the $m$-th trial, respectively.   In the simulation,  $M$ is set as $1000$.
All simulations are executed in MATLAB R2024b with an Intel Core Ultra 9 185H CPU @ 2.30 GHz.

Two examples are  designed   to verify the effectiveness of the proposed algorithm.  We refer to  our proposed  consensus-based distributed filter as CBDF.
Example 1 aims to demonstrate  the steady-state  performance of the proposed  filter as the fusion step number  $l$ increases. Example 2 is designed to  show the performances  and properties   in comparison with other existing distributed filers.

\begin{figure}[!htb]
	\centering
	{\includegraphics[width=2.4in]{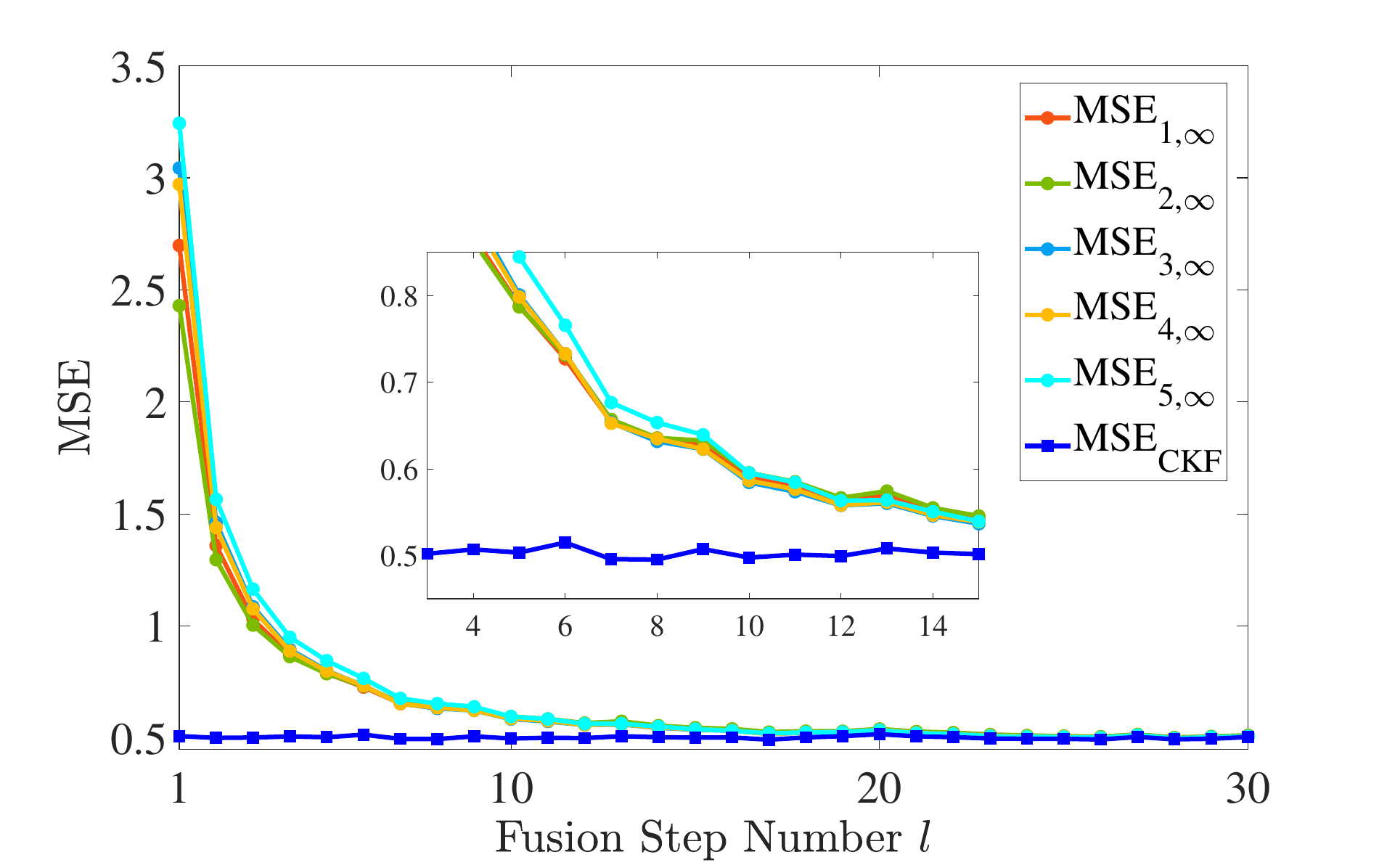}}
	\caption{Illustration figure for the steady-state performance of CBDF with the increasing fusion step number $l$.}
	\label{fig exampel_increaseL_directed}
\end{figure}

Example 1:
Fig. \ref{fig exampel_increaseL_directed}  exhibits  the steady-state performance of  CBDF of  five sensors with the increasing  fusion step number~$l$. It is shown that  the performance gap  between the centralized filter and  CBDF is  exponential convergence (Theorem~\ref{theorem exponential convergence}). In addition, a small  fusion
step can also ensure  the performance of CBDF.

\begin{figure}[!htb]
	\centering
	{\includegraphics[width=2.4in]{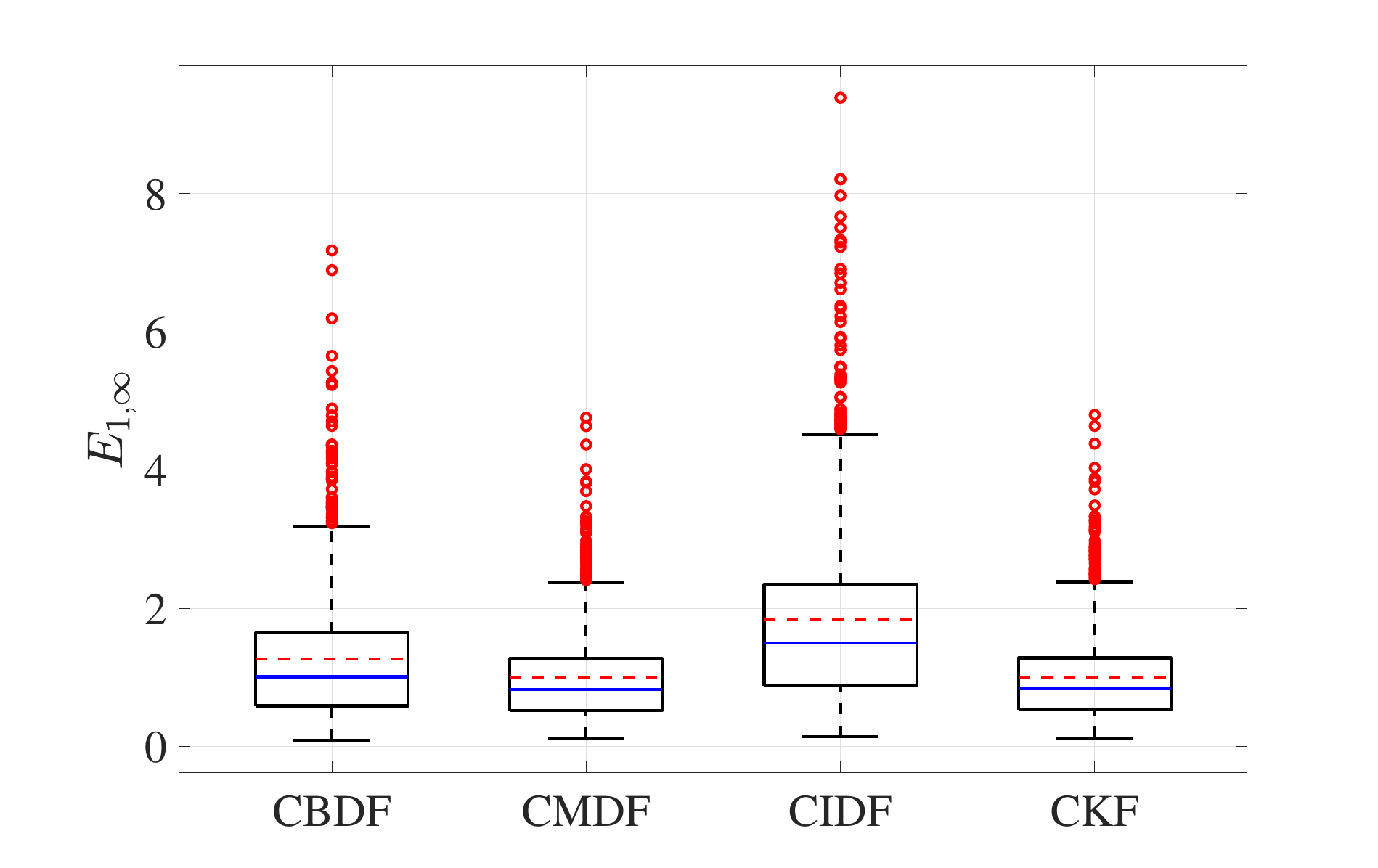}}
	\caption{The steady-state error of sensor~1 across 1000 Monte Carlo simulations  for four algorithms. In each box plot, the red dotted line represents the mean, while the blue solid line indicates the median. The upper and lower edges of each box correspond to the 25th and 75th percentiles, respectively. Red circles are used to highlight outliers that fall beyond 1.5 times the interquartile range.}
	\label{fig error_box}
\end{figure}

Example 2: To assess the performance of the proposed CBDF, three other algorithms are considered:
the consensus-on-information  distributed filter (CIDF) from \cite{battistelli2014kullback},
asymptotically optimal distributed filter (CMDF) from \cite{battistelli2014consensus},
and  the  centralized Kalman filter (CKF) in Section  \ref{sec centrailized estimator}. For comparison,  it is assumed that the communication topology is undirected  in Fig.~\ref{fig communication_topology simulation 1}, and the fusion step is set as $l=10$.
The Metropolis weights are adopted  in CIDF and CMDF.
%, and   the gain parameter is selected as 35 in AODF.
 We define  the  steady-state  error for  sensor~$i$ as
\begin{equation}
\begin{aligned}
E_{i,\infty} = \Vert \hat x_{i,\infty} - x_{\infty} \Vert^2_2.
\end{aligned}
\end{equation}
Fig.~\ref{fig error_box} shows  the steady-state error of sensor~1  for  four algorithms from  1000 Monte Carlo experiments. It is  observed  that CKF  exhibits  the best performance among the  four algorithms, while  CBDF demonstrates good performance. It is worth mentioning that  this simulation is executed  over undirected graphs for  comparison,   since the weight design methods  are  not  well  established  for  directed graphs \cite{rego2019distributed}.

Table~\ref{table time consumption}  presents the execution time  and memory usage of the four algorithms in Example~2.  All four algorithms exhibit similar   memory usage, but substantial differences emerge in terms of time consumption.  The CBDF  requires  less  computation time,  owing to its reliance primarily on simple addition and subtraction operations.

\begin{table}[!htb]
	\caption{The execution time and memory usage  in Example 2.}
	\label{table time consumption} % 标签中避免使用空格
	\centering
	\setlength{\tabcolsep}{2mm} % 减小列间距
	\renewcommand{\arraystretch}{1.2} % 减小行高
	\begin{tabular} {lcc}
		%{p{45pt}p{75pt}p{75pt}} % 使用更紧凑的列格式
		\hline
		Algorithm & Execution Time (s) & Memory Usage (MB) \\
		\hline
		CBDF & 5.21 & 3294 \\
		CMDF & 14.49 & 3341 \\
		CIDF & 41.69 & 3295 \\
		CKF & 1.07 & 3326 \\
		\hline
	\end{tabular}
\end{table}

\section{Conclusion}\label{sec conclusion}

This paper proposed a consensus-based  distributed filter over directed graphs, embedded with an augmented leader-following  information  fusion strategy. The parameters were  designed to guarantee the  uniform upper bound on the estimation error covariances. The steady-state  performance was  analyzed as  the  fusion step number increases, and the relations  between the proposed distributed filter  and the centralized filter were  revealed.
In the future,  we aim  to design a dynamic gain matrix to  optimize the distributed  filter at each  step, with the goal  of achieving   faster convergence. Additionally,  we intend   to explore  a new parameter design method to reduce  the spectral radius of the communication matrix.

\begin{appendices}

\section{PROOF of Proposition \ref{lemma measurement estimation error}}\label{app lemma measurement estimation error}

Item 1):
 Using equation (\ref{eq estimate measurement lsteo augmented}),  $\varepsilon^{(l)}_{i,k}$ can be derived as
\begin{equation*}
\begin{aligned}
\varepsilon^{(l)}_{i,k}& = z^{(l-1)}_{i,k}-\Lambda_i\Big[\sum^N_{l=1}a_{il}(z^{(l-1)}_{i,k}-z^{(l-1)}_{l,k})\\
&~~~~+ B_i(z^{(l-1)}_{i,k}-y_{k})\Big]-y_{k}\\
&=\varepsilon^{(l-1)}_{i,k} -\Lambda_i\Big[\sum^N_{l=1}a_{il}(\varepsilon^{(l-1)}_{i,k}-\varepsilon^{(l-1)}_{l,k})+ B_i\varepsilon^{(l-1)}_{i,k}\Big].
\end{aligned}
\end{equation*}
Then, the augmented vector $\varepsilon^{(l)}_{k}$ can be obtained as  $\varepsilon^{(l)}_{k} =G\varepsilon^{(l-1)}_{k}$.
When $l=0$, it follows  $\varepsilon^{(0)}_{i,k} =  z^{(0)}_{i,k}-y_{k}= CAe_{i,k-1}-C\omega_{k-1}-\nu_k.$
Similarly,    the augmented vector $\varepsilon^{(0)}_{k}$ is
$\varepsilon^{(0)}_{k}= (I_N\otimes CA)
e_{k-1}-(I_N\otimes C)(1_N\otimes\omega_{k-1})-1_N\otimes\nu_k.$
Next, it can be concluded that  $\varepsilon^{(l)}_{k}=G\varepsilon^{(l-1)}_{k}= G^{l}\varepsilon^{(0)}_{k}.$

Item 2):  Based on the results of Item 1),
$E\{\varepsilon^{(l)}_{k}\}$ and $E\{\varepsilon^{(l)}_{k}(\varepsilon^{(l)}_{k})^T\}$ can be calculated.

Item 3):   Utilizing the state estimator  (\ref{eq filter update}) in conjunction with  the  dynamical system (\ref{eq dynamics}), the estimation error $e_{i,k}$ for sensor~$i$  is
\begin{equation*}
\begin{aligned}
e_{i,k} &= (A-KCA) \hat x_{i,k-1|k-1} + Kz^{(l)}_{i,k} -Ax_{k-1}-\omega_{k-1}\\
& = (A-KCA) e_{i,k-1} + K\varepsilon^{(l)}_{i,k} -(I-KC)\omega_{k-1}+K\nu_k.
\end{aligned}
\end{equation*}
Subsequently,  by using  Proposition \ref{lemma measurement estimation error}, the  augmented estimation error  $e_k$, i.e.,  $e_{k} = [e^T_{1,k}, \ldots, e^T_{N,k}]^T$,  can be obtained as
\begin{equation}\label{eq lemma state error 1}
\begin{aligned}
e_{k} &= ( I_N \otimes (A-KCA)) e_{k-1} + (I_N \otimes K)G^l(I_N\otimes CA)e_{k-1}\\
&~~~~- (I_N \otimes K)G^l(1_N \otimes  (C\omega_{k-1}+\nu_k))\\
&~~~~+ 1_N \otimes (-(I-KC)\omega_{k-1}+K\nu_k).
\end{aligned}
\end{equation}
Note that $1_N \otimes (K\nu_k)=(I_N\otimes K)(1_N\otimes \nu_k)$
and $1_N \otimes  (C\omega_{k-1})= (I_N\otimes C)(1_N\otimes \omega_{k-1})$,  and  (\ref{eq lemma state error 1}) can be rewritten as
\begin{equation*}
\begin{aligned}
e_{k}
& = ( I_N \otimes(A-KCA)+(I_N \otimes K)G^l(I_N\otimes CA)) e_{k-1}\\
&~~~~-(I_N\otimes(I-KC)-(I_N \otimes K)\\
&~~~~\times G^l(I_N\otimes C))(1_N\otimes \omega_{k-1})\\
&~~~~+ (I_N \otimes K)(I_{Nr}-G^l)(1_N\otimes \nu_k).
\end{aligned}
\end{equation*}

By denoting the  matrices as  (\ref{eq lemma state estimation error A}),  (\ref{eq lemma state estimation error B}),  and    (\ref{eq lemma state estimation error D}),  $e_{k}$  has the following form  $e_{k} = \mathcal{A}(l) e_{k-1}-\mathcal{B}(l)(1_N\otimes \omega_{k-1})+\mathcal{D}(l)(1_N\otimes \nu_k).$

Item 4):  $P_{k|k}$ can be calculated as (\ref{eq lemma state estimation error P}) and  (\ref{eq lemma state estimation error phi}) according to $P_{k|k} = E\{e_ke^T_k\}$ .

\section{PROOF of Lemma \ref{lemma mu range}}\label{app lemma mu range}

Consider the row sum of the block matrix, and it holds
\begin{equation}\label{eq sum gijl}
\begin{aligned}
~~~\sum^n_{j=1}g_{[ij,l]} %&= \Big(1- \mu_{il}\sum^{N}_{j=1}{l_{ij}}-\mu_{il}a_{il}\Big)\otimes I_{r_l}\\
& = \Big(1 -\mu_{il}a_{il}\Big)\otimes I_{r_l}.
\end{aligned}
\end{equation}
 Under the condition $0<\mu_{il}\leq \frac{1}{l_{ii}+a_{il}}$ and the fact  $l_{ij}\leq 0, i\neq j$, it holds   $1-\mu_{il}(l_{ii}+a_{il})\geq 0$  and $-\mu_{il}l_{ij}\geq 0.$
By combining  (\ref{eq giil}) and  (\ref{eq gijl}),   it can be concluded that
the matrix $G$ is nonnegative.
By utilizing Lemma~\ref{lemma rho A by sum element}  and observing (\ref{eq sum gijl}), it follows
$\rho(G)\leq \text{max}_{1\leq i\leq n}\sum^n_{j=1}G_{ij} \leq \text{max}_{1\leq i\leq n} \sum^n_{j=1}g_{[ij,l]} \leq 1 -\mu_{il}a_{il}.$
Hence,  the conclusion is drawn that  $\rho(G)\leq 1$.
Subsequently,   it will be demonstrated  that the eigenvalues of $G$ do not include the value $1$.  By substituting the value $1$ into the characteristic polynomial of $G$, one has  $\vert I-G \vert = \vert \Lambda(\mathcal{L}\otimes I_r + B) \vert.$

According to Definition \ref{definition irreducibly diagnoally dominant},
$L_B = \mathcal{L}\otimes I_r +B$ is irreducibly diagonally dominant. Then, by utilizing Lemma~\ref{lemma domint matrix property},  $L_B = \mathcal{L}\otimes I_r +B$ is nonsingular.  Additionally, $\Lambda$ is positive definite with $\text{rank}(\Lambda) = Nr$, since $\mu_{ij}>0$. Utilizing  Sylvester's rank inequality, one obtains  $\text{rank}(\Lambda) + \text{rank}(L_B) - Nr\leq \text{rank}(\Lambda L_B)$ and $\text{rank}(\Lambda L_B) = Nr$. Hence, 1 is not an eigenvalue of the matrix  $G$. In conclusion, it can be inferred  that $ \rho(G)< 1$.

\section{PROOF of Theorem \ref{theo l0 l}}\label{app theo l0 l}

 By employing  Lemma~\ref{lemma Akto0} and considering  (\ref{eq lemma state estimation error A}),
it is evident  that    $\rho(\mathcal{A}(l))<1$ can guarantee the convergence of the distributed filter.   The following results provide a lower bound of the fusion  step number  $l$.

 Since $\Vert G \Vert_2<1$, one can conclude that $G^l$ is uniformly upper-bounded for any positive $l$.   Considering  (\ref{eq lemma state estimation error B}) and (\ref{eq lemma state estimation error D}),  both  $\mathcal{B}(l)$  and   $\mathcal{D}(l)$  are also uniformly upper-bounded  due to the boundedness of $K$ and $C$.  Likewise,
  it can be deduced that $\Phi(l)$ in (\ref{eq lemma state estimation error phi}) is uniformly upper-bounded  by using the boundedness of   $Q$ and $R$.
If $\mathcal{A}(l)$ is Schur stable,  it can be concluded   that $P_{k|k}$ is uniformly upper-bounded.

Given Assumption~\ref{ass observable},  it is well known that  the matrix  $A-KCA$ is Schur stable \cite{anderson2012optimal}. Applying  Lemma~\ref{lemma Akto0}, as $l$ tends to infinity, $\lim_{l\to \infty}G^l\to 0$. Consequently, one observes  $\lim_{l\to \infty}(I_N \otimes K)G^l(I_N\otimes CA)\to 0$. It is no doubt that there exists a positive $l$ such that $\mathcal{A}(l)$ is Schur stable.
In matrix theory, it is established that the spectral radius of
$\rho(\mathcal{A}(l))$ is bounded by the spectral norm $\Vert\mathcal{A}(l)\Vert_2$, i.e., $\rho(\mathcal{A}(l))\leq \Vert\mathcal{A}(l)\Vert_2$.
Using the identity $\Vert J_1\otimes J_2\Vert_2 = \Vert J_1\Vert_2\Vert J_2\Vert_2$, it follows  $\Vert\mathcal{A}(l)\Vert_2\leq\Vert A-KCA\Vert_2+\Vert K\Vert_2\Vert G\Vert^l_2\Vert CA\Vert_2.$
Let $\Vert A-KCA\Vert_2+\Vert K\Vert_2\Vert G\Vert^l_2\Vert CA\Vert_2< 1$. After  some algebraic manipulations,  one has
\begin{equation*}
\begin{aligned}
\Vert G\Vert^l_2<\frac{1-\Vert A-KCA\Vert_2}{\Vert K\Vert_2\Vert CA\Vert_2}.
\end{aligned}
\end{equation*}
Since  $0<\Vert G\Vert_2<1$, there exists  $l_0=\text{log}_{\Vert G\Vert_2}\frac{1-\Vert A-KCA\Vert_2}{\Vert K\Vert_2\Vert CA\Vert_2},$
such that when $l>l_0$, $\mathcal{A}(l)$ is Schur stable. Consequently, it can be proven that $P_{k|k}$ is uniformly upper-bounded.    Since
\begin{equation}\label{eq p_i  Pkk}
\begin{aligned}
P_{i,k|k} & = [0_{n\times n(i-1)},I_n,0_{n\times n(N-i)}]P_{k|k}\\
&~~~~\times[0^T_{n\times n(i-1)},I^T_n,0^T_{n\times n(N-i)}]^T,
\end{aligned}
\end{equation}
it can be found that $P_{i,k|k}$ is also uniformly upper-bounded.

\section{PROOF of Theorem \ref{theo converge to DLE}}\label{app theo converge to DLE}

Under Assumptions \ref{ass observable}, \ref{ass graph connected}, and \ref{ass G lees than 1},
it can be concluded that   $\mathcal{A}(l)$ and  $\mathcal{A}_{cc}$
are Schur stable.  In addition,  $\Phi(l)$  and $\Phi_{cc}$ are also uniformly upper-bounded due to the boundedness of $K$, $C$, $Q$, $R$, and $G^l$. By utilizing Theorem 1 in \cite{cattivelli2010diffusion}, it can be proven that  $P_{k|k}$  and $P_{cc,k|k}$   converge to  $P^{(l)}$ and $P_{cc}$, respectively.

\section{PROOF of Lemma \ref{lemma differnce series}}\label{app lemma differnce series}

First, consider the term  $\mathcal{A}(l)P^{(l)}\mathcal{A}^T(l)$  in  $ P^{(l)}$, and one has
\begin{equation}\label{eq lemma pl-pcc apa}
\begin{aligned}
\mathcal{A}(l)P^{(l)}\mathcal{A}^T(l)&=(\mathcal{A}_{cc}+ {\mathcal{\bar A}}(l))P^{(l)}(\mathcal{A}_{cc}+\bar {\mathcal{A}}(l))^T\\
& =  \mathcal{A}_{cc}P^{(l)} \mathcal{A}^T_{cc} + {\mathcal{\bar A}}(l)P^{(l)}\mathcal{A}^T_{cc}\\
&~~~+\mathcal{A}_{cc}P^{(l)} {\mathcal{\bar A}}^T(l)+{\mathcal{\bar A}}(l)P^{(l)}{\mathcal{\bar A}}^T(l).
\end{aligned}
\end{equation}
By utilizing (\ref{eq lemma pl-pcc apa}),  $P^{(l)}  - P_{cc}$ has the following form
\begin{equation}\label{eq lemma pl-pcc iteration}
\begin{aligned}
 P^{(l)}  - P_{cc} = \mathcal{A}_{cc}(P^{(l)}-P_{cc})\mathcal{A}^T_{cc}+ \bar \Phi(l),
\end{aligned}
\end{equation}
where
\begin{equation}\label{eq lemma pl-pcc bar phi}
\begin{aligned}
 \bar \Phi(l) & = {\mathcal{\bar A}}(l)P^{(l)}\mathcal{A}^T_{cc}+\mathcal{A}_{cc}P^{(l)} {\mathcal{\bar A}}^T(l)+{\mathcal{\bar A}}(l)P^{(l)}{\mathcal{\bar A}}^T(l)\\
 &~~~+\Phi(l)-\Phi_{cc}.
\end{aligned}
\end{equation}
Performing the iteration  (\ref{eq lemma pl-pcc iteration}) for $k$ times, it follows
\begin{equation}
\begin{aligned}
 P^{(l)}  - P_{cc}&= \mathcal{A}^k_{cc}(P^{(l)}-P_{cc})(\mathcal{A}^T_{cc})^k+ \sum^{k-1}_{j=0} \mathcal{A}^{j}_{cc} \bar \Phi(l) (\mathcal{A}^T_{cc})^j.
\end{aligned}
\end{equation}
By performing an infinite number of iterations and utilizing the fact that  $\mathcal{A}_{cc}$  is Schur stable,  it holds  $\lim_{k\to\infty}\mathcal{A}^k_{cc}(P^{(l)}-P_{cc})(\mathcal{A}^T_{cc})^k=0$
 and  $ P^{(l)}  - P_{cc} = \sum^{\infty}_{j=0} \mathcal{A}^{j}_{cc} \bar \Phi(l) (\mathcal{A}^T_{cc})^j.$
Next, similarly to (\ref{eq lemma pl-pcc apa}),  $\bar \Phi(l)$ in  (\ref{eq lemma pl-pcc bar phi})
can be  calculated as (\ref{eq lemma bar phi fin}).

\section{PROOF of Theorem \ref{theorem exponential convergence}}\label{app theorem exponential convergence}

By calculating the spectral norm of  (\ref{eq lemma pl-pcc formal}), one has
\begin{equation}\label{eq pl-pcc2}
\begin{aligned}
\Vert P^{(l)}  - P_{cc}\Vert_2 \leq \Vert \bar \Phi(l) \Vert_2 \sum^{\infty}_{k=0} \Vert \mathcal{A}^k_{cc} \Vert^2_2.
\end{aligned}
\end{equation}
For $\Vert \bar \Phi(l) \Vert_2$,  based on  (\ref{eq lemma bar phi fin}), it follows
\begin{equation}
\begin{aligned}
\Vert \bar \Phi(l)\Vert_2 & \leq (\Vert{\mathcal{\bar A}}(l)\Vert_2 +2\Vert\mathcal{A}_{cc}\Vert_2) \Vert P^{(l)}\Vert_2 \Vert{\mathcal{\bar A}}(l)\Vert_2\\
 &~~~+N(\Vert{\mathcal{\bar B}}(l)\Vert_2 +2\Vert\mathcal{B}_{cc}\Vert_2) \Vert Q\Vert_2 \Vert{\mathcal{\bar B}}(l)\Vert_2\\
 & ~~~+N(\Vert{\mathcal{\bar D}}(l)\Vert_2 +2\Vert\mathcal{D}_{cc}\Vert_2) \Vert R\Vert_2 \Vert{\mathcal{\bar D}}(l)\Vert_2.
\end{aligned}
\end{equation}
 By  conducting some calculations and  isolating   $\Vert G^l\Vert_2$   from $\Vert{\mathcal{\bar A}}(l)\Vert_2$, $\Vert{\mathcal{\bar B}}(l)\Vert_2$, and $\Vert{\mathcal{\bar D}}(l)\Vert_2$,     there exist a positive number $M_3$  such that
\begin{equation}\label{eq bar phi bound}
\begin{aligned}
\Vert \bar \Phi(l)\Vert_2\leq M_3 \Vert G^l\Vert_2
\end{aligned}
\end{equation}
holds, where
\begin{equation}
\begin{aligned}
M_3 &=  (\Vert{\mathcal{\bar A}}(l)\Vert_2 +2\Vert\mathcal{A}_{cc}\Vert_2) \Vert P^{(l)}\Vert_2 \Vert K\Vert_2\Vert CA\Vert_2\\
 &~~~+N(\Vert{\mathcal{\bar B}}(l)\Vert_2 +2\Vert\mathcal{B}_{cc}\Vert_2) \Vert Q\Vert_2 \Vert K\Vert_2\Vert C\Vert_2\\
 & ~~~+N(\Vert{\mathcal{\bar D}}(l)\Vert_2 +2\Vert\mathcal{D}_{cc}\Vert_2) \Vert R\Vert_2 \Vert K\Vert_2.
\end{aligned}
\end{equation}
By using Lemma \ref{lemma norm of Ak} and the similar technique in \cite{qian2022consensus}, a positive number $M_4$ can be found such that
\begin{equation}\label{eq A2 series}
\begin{aligned}
\Vert \mathcal{A}^k_{cc}\Vert_2 &\leq \sqrt{n}\sum^{n-1}_{j=0}\binom{n-1}{j}\binom{k}{j}\Vert \mathcal{A}_{cc}\Vert^j_2 \rho(\mathcal{A}_{cc})^{k-j}\\
& \leq \sqrt{n} \Bigg(n\text{max}_{j}\binom{n-1}{j}\Vert \mathcal{A}_{cc}\Vert^j_2\Bigg)k^n\rho(\mathcal{A}_{cc})^{k-n}\\
&\leq M_4k^n\rho(\mathcal{A}_{cc})^{k-n}.
\end{aligned}
\end{equation}
Since $\rho(\mathcal{A}_{cc}) < 1$, the convergence of the infinite sum  $\sum^{\infty}_{k=0} \Vert \mathcal{A}^k_{cc} \Vert_2$  can be proven. Since  $\Vert \mathcal{A}^k_{cc}\Vert_2$ is uniformly bounded for all $k$,  there exists a positive  number  $M_5$ such that
\begin{equation}\label{eq sum Acc leq M}
\begin{aligned}
\sum^{\infty}_{k=0} \Vert \mathcal{A}^k_{cc} \Vert^2_2 \leq M_5.
\end{aligned}
\end{equation}
Next, the term $ \Vert G^l\Vert_2$ is analyzed, and the results of two items are proven respectively.

Item 1): Similarly to (\ref{eq A2 series}),   there exists a positive number $M_6$ such that
\begin{equation}\label{eq Gl2 bound 1}
\begin{aligned}
\Vert G^l\Vert_2 &\leq \sqrt{Nr}\sum^{Nr-1}_{j=0}\binom{Nr-1}{j}\binom{l}{j}\Vert G\Vert^j_2 \rho(G)^{l-j}\\
& \leq M_6 l^{Nr}\rho(G)^{l-Nr}.
\end{aligned}
\end{equation}
By combining  (\ref{eq pl-pcc2}),   (\ref{eq bar phi bound}), (\ref{eq sum Acc leq M}), and (\ref{eq Gl2 bound 1}), a  positive number $M_1$ can be designed such that  $\Vert P^{(l)}  - P_{cc}\Vert_2\leq M_1 l^{Nr}\rho(G)^{l-Nr}.$

Item 2): If  $ \Vert G\Vert_2 <1$, (\ref{eq bar phi bound}) can be  calculated as
\begin{equation}\label{eq Gl2 bound 2}
\begin{aligned}
\Vert \bar \Phi(l)\Vert_2\leq M_3 \Vert G\Vert^l_2.
\end{aligned}
\end{equation}
By combining  (\ref{eq pl-pcc2}),   (\ref{eq sum Acc leq M}), and (\ref{eq Gl2 bound 2}),  a positive number $M_2$ can be obtained such that $\Vert P^{(l)}  - P_{cc}\Vert_2 \leq M_2 \Vert G\Vert^l_2.$

\end{appendices}

\bibliographystyle{IEEEtran}
\bibliography{ref}

\end{document}